\documentclass[twocolumn,
prl,
showpacs,
superscriptaddress,
floatfix
]{revtex4-1} 

\usepackage[dvips]{graphicx}
\usepackage{bm}
\usepackage{ulem}
\usepackage{amsmath}
\usepackage{amssymb}
\usepackage{color}

\usepackage[dvipdfmx,
colorlinks=true,
citecolor=blue,
setpagesize=false]{hyperref}

\begin{document}

\title{
From a quasimolecular band insulator to a relativistic Mott insulator in $t_{2g}^5$ systems with 
a honeycomb lattice structure 
}

\author {Beom Hyun Kim}
\affiliation{
  Computational Condensed Matter Physics Laboratory, 
  RIKEN, Wako, Saitama 351-0198, Japan}
\affiliation{
  Interdisciplinary Theoretical Science (iTHES) Research Group,
  RIKEN, Wako, Saitama 351-0198, Japan}
\author {Tomonori Shirakawa}
\affiliation{Computational Quantum Matter Research Team, 
   RIKEN Center for Emergent Matter Science (CEMS), 
   Wako, Saitama 351-0198, Japan}
\author {Seiji Yunoki}
\affiliation{
  Computational Condensed Matter Physics Laboratory, 
  RIKEN, Wako, Saitama 351-0198, Japan}
\affiliation{
  Interdisciplinary Theoretical Science (iTHES) Research Group,
  RIKEN, Wako, Saitama 351-0198, Japan}
\affiliation{Computational Quantum Matter Research Team, 
   RIKEN Center for Emergent Matter Science (CEMS), 
   Wako, Saitama 351-0198, Japan}
\affiliation{
  Computational Materials Science Research Team, 
  RIKEN Advanced Institute for Computational Science (AICS),  
  Kobe, Hyogo 650-0047, Japan}

\date{\today}

\begin{abstract}
The $t_{2g}$ orbitals of an edge-shared transition-metal oxide with a honeycomb lattice structure 
form dispersionless electronic bands when only hopping mediated by the edge-sharing
oxygens is accessible. 
This is due to the formation of isolated quasimolecular orbitals (QMOs) in each hexagon, 
introduced recently by Mazin \textit{et al.} 
[\href{http://journals.aps.org/prl/abstract/10.1103/PhysRevLett.109.197201}
{Phys. Rev. Lett. \textbf{109}, 197201 (2012)}], 
which stabilizes a band insulating phase 
for $t_{2g}^5$ systems. 
However, with help of the exact diagonalization method to treat the electron kinetics and correlations 
on an equal footing, we find that 
the QMOs are fragile against not only the spin-orbit coupling (SOC) but also  
the Coulomb repulsion.
We show that the electronic phase of $t_{2g}^5$ systems 
can vary from 
a quasimolecular band insulator to a relativistic $J_{\rm eff}=1/2$ Mott insulator 
with increasing the SOC as well as the Coulomb repulsion. 
The different electronic phases manifest themselves 
in electronic excitations observed in optical conductivity and resonant 
inelastic x-ray scattering. 
Based on our calculations, we assert that 
the currently known Ru$^{3+}$- and Ir$^{4+}$-based honeycomb systems 
are far from the quasimolecular band insulator but rather the relativistic Mott insulator. 
\end{abstract}


\maketitle


{\it Introduction ---} 
Physical properties of $4d$ and $5d$ transition metal (TM) compounds with 
nominally less than six $d$ electrons are 
determined by the $t_{2g}$ manifold because of a strong cubic crystal field.
A strong spin-orbit coupling (SOC) causes $t_{2g}$ orbitals
split into effective total angular momenta 
$j_{\rm eff}=1/2$ and $3/2$.
The relativistic electronic feature in these TM compounds has drawn much attraction recently 
as exotic electronic, magnetic, and topological phases have been expected, 
including $J_{\rm eff}=1/2$ Mott insulator~\cite{BJKim2008,BJKim2009,Jackeli2009,Watanabe2010,Martins2011}, 
superconductivity~\cite{Wang2011,Watanabe2013},
topological insulator~\cite{Pesin2010,Yang2010}, 
Weyl semimetal~\cite{Krempa2012,AGo2012}, 
and spin liquid~\cite{Okamoto2007,Singh2013}.
Among them, the research on $t_{2g}^5$ systems 
forming a honeycomb lattice structure 
with edge-sharing ligands 
has been triggered by 
the possibility of a nontrivial topological phase~\cite{Shitade2009,CHKim2012} 
or a Kitaev-type spin liquid~\cite{Chaloupka2010,Jiang2011,Reuther2011}, 
attributed to their unique hopping geometry.
However, the existing compounds such as Na$_2$IrO$_3$~\cite{Singh2010,Liu2011,Choi2012,
Ye2012}, Li$_2$IrO$_3$~\cite{Singh2012}, Li$_2$RhO$_3$~\cite{Luo2013},
and $\alpha$-RuCl$_3$~\cite{Sears2015,Majumder2015,Johnson2015}, 
have been turned out to be magnetic insulators 
with a long-range antiferromagnetic (AFM) or glassy-spin order.

In order to understand the electronic and magnetic structures of these compounds with 
$t_{2g}^5$ configuration, two distinct points of view, i.e., Mott- and Slater-type pictures, have been proposed. 
In the Mott picture, the Coulomb repulsion opens the gap of 
the relativistic $j_{\rm eff}=1/2$ based band 
and the superexchange interaction between the relativistic isospins 
stabilizes the AFM order~\cite{Comin2012,Gretarsson2013,Sohn2013,BHKim2014,
Plumb2014,HSKim2015}. 
This strong coupling approach can successfully elucidate the observed excitations 
in the optical conductivity (OC) and resonant inelastic x-ray scattering (RIXS) 
for Na$_2$IrO$_3$~\cite{BHKim2014}.
However, there has been still debate on the origin of the zigzag AFM order 
in Na$_2$IrO$_3$~\cite{Kimchi2011,Chaloupka2013}.
In contrast, the Slater picture focuses on the itinerant nature of $t_{2g}$ bands and treats 
the Coulomb interactions perturbatively. This weak coupling approach 
naturally explains the zigzag AFM order with a concomitantly induced band 
gap~\cite{Mazin2012,Mazin2013,Foyevtsova2013,HJKim2014}. 
However, it is difficult to fully describe the observed excitations in the OC and RIXS 
for Na$_2$IrO$_3$~\cite{Li2015,MJKim2016,Igarashi2015}. 
Because 
the hopping integral, Coulomb repulsion, and SOC 
are of similar energy scale, either of these two opposite pictures cannot be ruled out. 

As the electron hoppings between the adjacent TMs via the two edge-sharing 
ligands are highly orbital dependent~\cite{Supp},
the electron motion is confined within a 
single hexagon formed by six TMs. 
This has led to the notion of the quasimolecular orbital (QMO) formation~\cite{Mazin2012},
where 
each $t_{2g}$ orbital of a TM participates in the formation of QMO at one of the three 
different hexagons around the TM. 
Therefore, when the SOC and Coulomb repulsion are both small, the ground state 
for $t_{2g}^5$ systems is a band insulator with a strong QMO character. 
However, when the SOC is strong, the QMOs are no longer well defined
because the SOC induces an effective hopping between neighboring QMOs. 
In this limit, the local relativistic 
$j_{\rm eff}$ orbitals are instead expected to play a dominant role in characterizing 
the electronic and magnetic structures. 

In this paper, by considering a minimal microscopic model which  
captures both extreme limits, 
we examine the ground state phase diagram for $t_{2g}^5$ electron configuration 
with help of the numerically exact diagonalization method. 
We show that not only the SOC but also the Coulomb repulsion induces a crossover 
of the ground state with the strong QMO to 
relativistic $j_{\rm eff}$ orbital character. 
Concomitantly, the nature of the emerging electronic state varies 
from a quasimolecular band insulator to a relativistic $J_{\rm eff}=1/2$ Mott insulator. 
The different electronic states are manifested in 
distinct behaviors of excitations, directly observed in 
OC and RIXS experiments. 
Our analysis concludes that the currently known Ru$^{3+}$- and Ir$^{4+}$-based systems 
are both far from the QMO state but rather the relativistic Mott insulator.

\begin{figure}[t]
\centering
\includegraphics[width=.9\columnwidth]{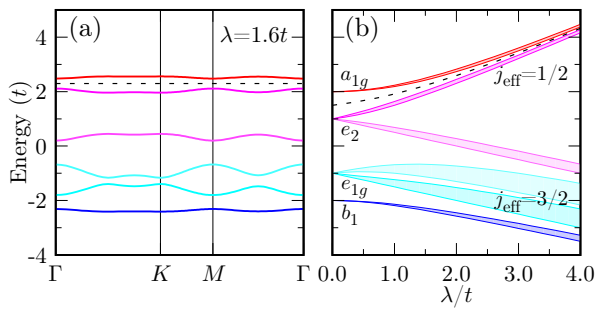}
\caption
{ (Color online)
(a) Noninteracting band dispersion with the hopping $t\, (>0)$ mediated by
edge-sharing ligands for the SOC $\lambda=1.6 t$. 
Each band is doubly degenerate (Kramer's doublet). 
(b) Energy splitting of the noninteracting bands as a function of $\lambda$. 
Shade region represents the band width of each band. 
$a_{1g}$, $e_{2u}$, $e_{1g}$, and $b_{1u}$ refer to the symmetries of the quasimolecular bands
when $\lambda=0$.
Dotted lines in (a) and (b) denote the Fermi energy for $t_{2g}^5$ systems. 
}
\label{fig1}
\end{figure}

{\it Noninteracting limit --}
Without the SOC, 
the QMOs are the exact eigenstates with $b_{1u}$, $e_{1g}$, $e_{2u}$, and $a_{1g}$ 
symmetries, the eigenenergies being $-2t$, $-t$, $t$, and $2t$ 
($t$: hopping between adjacent TMs), respectively~\cite{Supp}, 
and they form dispersionless bands. 
However, as shown in Fig.~\ref{fig1}(a), once the SOC $\lambda$ is finite, 
the double degeneracy of $e_{1g}$ and $e_{2u}$ symmetries is 
lift and the QMOs are split in total into six Kramer's doublet bands with
finite dispersion. 
With further increasing $\lambda$, the highest two bands as well as the lowest four bands 
come close in energy but the energy splitting between these two manifolds becomes larger, 
smoothly connecting to the $j_{\rm eff}=1/2$ and $3/2$ based bands 
[see Fig.~\ref{fig1}(b)]. 
Thus, as $\lambda$ increases, an insulating gap of $t_{2g}^5$ systems for $t>0$ 
gradually decreases with continuous change of hole character 
from $a_{1g}$ to $j_{\textrm{eff}}=1/2$.

\begin{figure}[tb]
\centering
\includegraphics[width=.99\columnwidth]{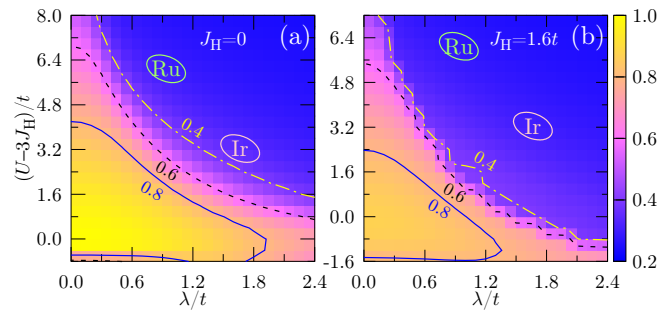}
\caption
{ (Color online)
  The hole density $\bar{n}_{a_{1g}}$ of the $a_{1g}$ quasimolecular band at the $\Gamma$ point 
  as functions of $U$ and $\lambda$ for (a) $J_{\rm H}=0$ and (b) $J_{\rm H}=1.6t$.
Note that $\bar{n}_{a_{1g}}=1$ for the pure QMO state. 
The loci of Ru$^{3+}$- and Ir$^{4+}$-based $4d$ and $5d$ transition-metal compounds are 
also indicated. 
}
\label{fig2}
\end{figure}

{\it Correlation effect ---}
Let us now explore the effect of electron correlations by considering a 
three-band Hubbard model on a 
periodic six-site cluster for electron density $n=5$~\cite{Supp} 
with Lanczos exact diagonalization method~\cite{Morgan93,Dagotto94}, 
which allows us to treat the electron kinetics inducing the QMO formation and the electron 
correlations on an equal footing, thus clearly going beyond the previous study~\cite{BHKim2014}. 
We first examine the ground state 
as functions of the intra-orbital Coulomb repulsion $U$, Hund's coupling $J_{\rm H}$, and $\lambda$, 
and check the stability of the QMO state. 
For this purpose, we calculate the hole density $\bar{n}_{a_{1g}}$ 
of the $a_{1g}$ quasimolecular band at the $\Gamma$ point, 
which is exactly one for the pure QMO state~\cite{Supp}. 
Figure~\ref{fig2}(a) shows the result of $\bar{n}_{a_{1g}}$ 
for $J_{\rm H}=0$ with varying $U$ and $\lambda$. 
It clearly demonstrates that $\lambda$ and $U$ are both destructive perturbations to the QMOs.
As already pointed out in Ref.~\cite{Foyevtsova2013},
the strong SOC mixes the three $t_{2g}$ orbitals at each site, 
which gives rise to a finite overlap between the QMOs in neighboring hexagons, 
thus unfavorable to the QMO formation. 
More interestingly, we find here in Fig.~\ref{fig2}(a) that the
Coulomb repulsion is also adequate 
to destroy the QMO state. This is understood because the Coulomb interactions promote 
the scattering among electrons bounded in adjacent hexagons.

The ground state is described by a direct product of local states with not only $d^5$ electron configuration 
but also other configurations such as $d^4$ and $d^6$. Therefore, the ground state is sensitive to the local multiplet 
structures of these electron configurations.  
According to the multiplet theory, 
the Hund's coupling $J_{\rm H}$ always brings about additional splitting of the $d^4$ multiplet
hierarchy~\cite{Supp}.
It is thus easily conjectured that $J_{\rm H}$ also plays a role in the QMO formation. 
Figure~\ref{fig2}(b) well represents the effect of $J_{\rm H}$ on the QMO state. 
In finite $J_{\rm H}$, the region with the strong QMO character shrinks with somewhat smaller 
$\bar{n}_{a_{1g}}$ 
and the crossover boundary becomes sharper.

\begin{figure}[tb]
\centering
\includegraphics[width=.95\columnwidth]{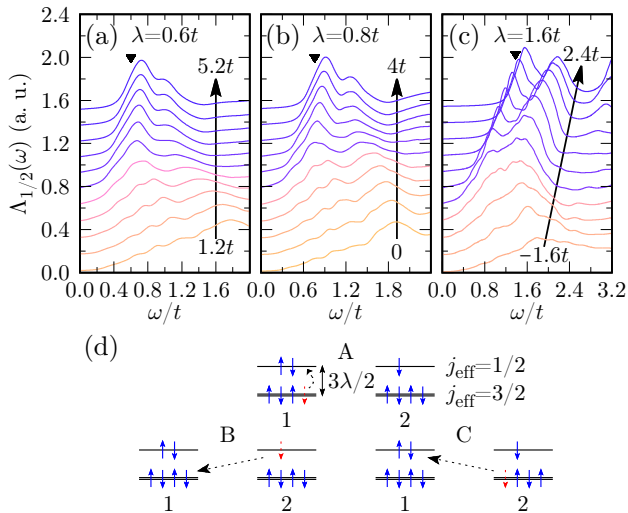}
\caption
{ (Color online)
 (a)--(c): Excitation spectrum $\Lambda_{1/2}(\omega)$ of 
 a doublon-holon pair formed in the neighboring $j_{\rm eff}=1/2$ orbitals
 for (a) $\lambda=0.6t$, (b) $0.8t$, and (c) $1.6t$ with different  
 $U-3J_{\rm H}$ values. 
 $U-3J_{\rm H}$ varies from (a) $1.2t$ to $5.2t$, (b) $0$ to $4t$, 
 and (c) $-1.6t$ to $2.4t$ with the same increment of $0.4t$.
 Solid triangles indicate peak positions 
 of the exciton-like excitations in the RIXS spectrum 
 for $U-3J_{\rm H}=5.2t$ (a), $4t$ (b), and $2t$ (c). 
 We set $J_{\rm H}=1.6t$ and 
 different line colors correspond to the values of $\bar{n}_{a_{1g}}$ shown 
 in Fig.~\ref{fig2}(b). 
 (d) Schematic energy diagrams of three relevant electron-hole excitations. 
  Blue solid and red dotted arrows refer to electron and hole, respectively, 
  and up and down arrows denote Kramer's doublet with positive (up arrow) and negative 
  (down arrow) eigenvalues of $j^z_{\rm eff}$.
 ``1" and ``2" indicate two neighboring TM sites. 
}
\label{fig3}
\end{figure}

In the whole parameter region of Fig.~\ref{fig2}, the ground state is insulating. 
However, the insulating nature is expected to vary
from a band insulator to a Mott insulator across the
boundary where the QMO character is abruptly lost. 
To verify this conjecture, we calculate the excitation spectrum $\Lambda_{1/2}(\omega)$ of 
a doublon-holon pair formed in the $j_{\rm eff}=1/2$ orbitals at neighboring sites~\cite{Supp}, 
excitations schematically shown in `B' of Fig.~\ref{fig3}(d), 
which directly reflects the  charge gap structure.  
As shown in Figs.~\ref{fig3}(a)--(c), 
the $U$ dependence of $\Lambda_{1/2}(\omega)$ is 
qualitatively different  
across the crossover boundary.
Below the boundary where the QMO character is strong,
the low energy excitations shift downward in spite of increasing $U$, 
implying that an insulating gap evidently decreases with increasing the electron repulsion.
In contrast, above the boundary where the QMO character is lost, the clear increase 
of the lowest peak position 
is manifested with increasing $U$, indicating the increase of the gap as in a Mott insulator. 
The similar feature is also found when $\lambda$ is increased with fixed $U$ and 
$J_{\rm H}$~\cite{Supp}.
These results support the conjecture that the insulating nature changes across 
the crossover boundary. 
The conjecture is further supported by other excitation spectra shown below.

\begin{figure}[tb]
\centering
\includegraphics[width=.95\columnwidth]{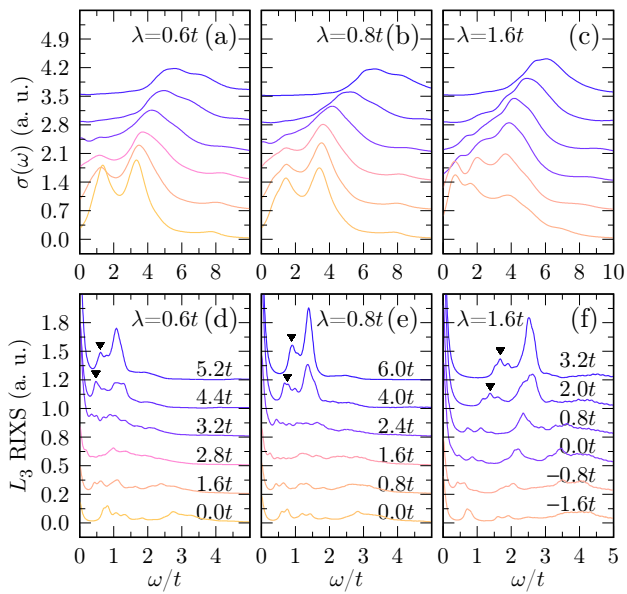}
\caption
{ (Color online)
 (a)--(c) OC $\sigma(\omega)$ and (d)--(f) $L_3$-edge RIXS spectrum 
  at wave number $\mathbf{q}=0$ 
  for $\lambda=0.6t$ [(a) and (d)], $0.8t$ [(b) and (e)], 
  and $1.6t$ [(c) and (f)]
  with various $U-3J_{\rm H}$ values indicated in the figures. 
  We set $J_{\rm H}=1.6t$ and 
  different line colors correspond to the values of $\bar{n}_{a_{1g}}$ shown in Fig.~\ref{fig2}(b).
  The exciton-like excitations in the RIXS spectrum are indicated by solid triangles in 
  (d)--(f).
}
\label{fig4}
\end{figure}

{\it OC and RIXS spectra ---}
The Kubo formula and the continued fraction method are exploited to 
investigate the OC and $L_3$-edge RIXS spectra~\cite{Supp}. 
Figure~\ref{fig4} summarizes the results 
for $\lambda=0.6t$, 0.8$t$, and 1.6$t$ with $J_{\rm H}$=$1.6t$. 
Note that the QMO character suddenly diminishes at 
$U-3J_{\rm H}\sim3t$ for $\lambda=0.6t$, $2t$ for $0.8t$, and $0$ for $1.6t$, 
as shown in Fig.~\ref{fig2}(b).
The abrupt change of the electronic characteristics is 
reflected in these excitation spectra.

When the QMO character is strong with large $\bar{n}_{a_{1g}}$, the OC exhibits 
a two-peak structure for $\lambda=0.6t$ and $0.8t$, and a three-peak structure 
for $\lambda=1.6t$.
The lower (higher) peak in the OC is 
attributed to the transition between the $a_{1g}$ and $e_{2u}$ ($e_{1g}$) quasimolecular bands 
whose energies are around $t$ ($3t$) (see Fig.~\ref{fig1}). 
The large splitting of the $e_{2u}$ bands for the strong SOC 
can give an additional splitting to the lower peak in the OC. 
The RIXS spectrum also shows the similar peak structures near the similar excitation energies. 
Consequently, the excitations can be interpreted on the basis of the single-particle picture.
Therefore, a strong band insulating character is predominant in this region.

In the region where the QMO character is degraded, the OC shows 
a one-peak structure and the peak position 
monotonically increases with $U$, while 
the RIXS spectra exhibits the dominant peak around $3\lambda/2$ 
in addition to an almost zero energy
peak due to the magnetic excitation.
These features can be well understood as on-site or inter-site electron-hole excitations 
in the local relativistic $j_{\rm eff}$ orbitals.
As shown in `B' and `C' of Fig.~\ref{fig3}(d), two types of inter-site electron-hole excitations 
can play a role in the OC.
However, the edge-shared geometry suppresses the contribution of type `B' electron-hole excitation 
simply because the hopping between the neighboring $j_{\rm eff}=1/2$ orbitals is zero~\cite{hopJ}. 
Hence, only type `C' electron-hole excitation gives the dominant contribution to form the one-peak like structure 
at excitation energy $\omega\approx U-3J_{\rm H}+\frac{3\lambda}{2}$. 
In the RIXS spectrum,
an on-site electron-hole excitation indicated in `A' of Fig.~\ref{fig3}(d), i.e., 
a local $d$-$d$ transition between the $j_{\rm eff}=3/2$ and $1/2$ orbitals, 
can give a dominant intensity at $\omega\approx 3\lambda/2$.
Thus, both OC and RIXS spectra in this region can be interpreted
in terms of the relativistic Mott insulating picture.

In the relativistic Mott insulating limit, the RIXS spectrum 
shows an additional peak below the local $d$-$d$ excitation ($\approx3\lambda/2$) and above 
the almost zero energy magnetic peak, 
which is marked by triangles in Figs.~\ref{fig4}(d)--(f).
Indeed, this peak has been observed in Na$_2$IrO$_3$ 
and the origin is attributed to the exciton formation induced by 
the inter-site electron correlations~\cite{Gretarsson2013}. 
However, the consecutive theoretical study based on the strong coupling model 
calculations has shown 
that this exciton-like peak appears even without considering the inter-site electron correlations 
when the inter-site migration of electrons from the $j_{\rm eff}=1/2$ to $3/2$ orbitals
[`B' of Fig.~\ref{fig3}(d)] follows the local $d$-$d$ transition [`A' of Fig.~\ref{fig3}(d)], 
resulting in the inter-site electron-hole excitation 
between the $j_{\rm eff}=1/2$ orbitals~\cite{BHKim2014}. 
As shown in Figs.~\ref{fig3}(a)--(c), the excitation spectrum $\Lambda_{1/2}(\omega)$ of 
a doublon-holon pair in the $j_{\rm eff}=1/2$ orbitals also yield the obvious spectral weight 
in the vicinity of the exciton-like RIXS excitation energy, implying that these excitations 
are due to the same origin.  
In addition, the monotonic increase of the exciton-like peak position in the RIXS spectrum with $U$ 
is also consistent with $\Lambda_{1/2}(\omega)$. 
We also find that the intensity of the exciton-like 
RIXS peak depends strongly on momentum~\cite{Supp},
which is in good
agreement with the experiments~\cite{Gretarsson2013}. 
Thus, it is reasonable to infer that the origin of the exciton-like peak near the edge of local $d$-$d$ 
excitation in the RIXS is due to 
the combined excitations of two different types (`A' and `B').

{\it Ir$^{4+}$- and Ru$^{3+}$-based systems ---} 
A typical SOC is known to be $0.4$--$0.5$ eV for $5d$ systems 
and $0.1$--$0.2$ eV for $4d$ systems~\cite{Dai2008,Clancy2012}.
Recent theoretical studies have estimated that $t\approx0.27$ eV for the most studied Ir$^{4+}$ 
system Na$_2$IrO$_3$~\cite{Mazin2012,Foyevtsova2013,Yamaji2014}, 
and $t\approx 0.11$ eV~\cite{HSKim2015} and $0.16$ eV~\cite{Winter2016} 
for $\alpha$-RuCl$_3$ (Ru$^{3+}$), both of which are much smaller than that for Ir$^{4+}$ systems. 
$U$ and $J_{\rm H}$, however, are not easy to be determined because
the full screening effect of electron correlations is hardly treated.
One useful expedient is to extract them from the OC measurement.

Since 
the dominant optical peak appears near $U-3J_{\rm H}+\frac{3\lambda}{2}$,
we can estimate $U-3J_{\rm H} \sim 0.8$--$1.0$ eV for Na$_2$IrO$_3$
based on the existing OC data, which exhibits a one-peak structure 
around 1.6 eV~\cite{Comin2012,Sohn2013}. 
Assuming $U-3J_{\rm H}=3.2t$ ($\approx 0.86$ eV), 
our results for the OC and RIXS spectra in Figs.~\ref{fig4}(c) 
and (f) are both indeed in good quantitative agreement with the 
experiments~\cite{Comin2012,Sohn2013,Gretarsson2013}, 
except for a double-peak structure around 0.7 eV ($\approx 2.6t$) 
observed in the RIXS experiment, instead of a single peak found in Fig.~\ref{fig4}(f). 
Here, our calculations assume the cubic crystal field.
However, 
the crystal structure of Na$_2$IrO$_3$ is known to display the additional trigonal 
distortion~\cite{note}, thus significantly departing from the 
ideal IrO$_6$ octahedra~\cite{Choi2012}. 
This additional distortion can mix the relativistic 
$j_{\rm eff}=1/2$ and $3/2$ orbitals, and lead to the splitting of
the dominant RIXS peak of the local $d$-$d$ transition into multiple subpeaks~\cite{Liu2012,Plotnikova2016}. 
Indeed, we find that the single peak found in Fig.~\ref{fig4}(f) is split into two 
(or multiple) subpeaks in the presence of the trigonal distortion~\cite{Supp} 
(see also Refs.~\cite{Gretarsson2013,BHKim2014}). 
Therefore, we can conclude that the electronic state of Na$_2$IrO$_3$
is far from the QMO limit and 
is located in the relativistic Mott insulating region, as indicated in Fig.~\ref{fig2}.

The recent optical absorption experiments for $\alpha$-RuCl$_3$
have found a dominant peak around 1.2 eV as well as a small additional 
peak near 0.3 eV~\cite{Plumb2014,Sandilands2016}.
The photoemission spectroscopy measurement has also observed a large 
gap about 1.2 eV~\cite{Koitzsch2016}.
Thus, $U-3J_{\rm H}$ for $\alpha$-RuCl$_3$ is estimated to be about 0.9--1.1 eV. 
Adopting $t=0.16$ eV,
our result for $U-3J_{\rm H}=6t$ ($\approx0.96$ eV) and $\lambda=0.8t$ ($\approx0.13$ eV)
in Fig.~\ref{fig4}(b)
also exhibits the dominant peak near 1.1 eV ($\approx6.9t$).
Although no experimental RIXS spectrum has been reported yet,
the recent neutron scattering measurement on $\alpha$-RuCl$_3$ observed an inelastic
peak near 195 meV and estimated that $\lambda \approx 130$ meV~\cite{Banerjee2016}.
This observation is consistent with a dominant RIXS peak around $1.4t$ 
($\approx0.22$ eV) for $U-3J_{\rm H}=6t$ in Fig.~\ref{fig4}(e).
Therefore, we expect that $\alpha$-RuCl$_3$ 
is located also in the QMO poor region, as indicated in Fig.~\ref{fig2}.

It should be noted, however, that the OC 
for $U-3J_{\rm H}=6t$ 
in Fig.~\ref{fig4}(b) fails to yield
the low-energy peak near 0.3 eV observed experimentally in $\alpha$-RuCl$_3$. 
This can be explained by considering an additional direct $d$-$d$ hopping $t'$ 
between neighboring sites, estimated as large as $-0.23$ eV in Ref.~\cite{HSKim2015} 
and $-0.15$ eV in Ref.~\cite{Winter2016}. 
As discussed in Supplemental Material~\cite{Supp}, the low-energy peak attributed to 
the local $d$-$d$ transition [`A' in Fig.~\ref{fig3}(d)] can emerge because the forbidden optical transition 
among the $j_{\rm eff}=1/2$ bands without the direct hopping $t'$ can now 
be accessed in the presence of $t'$~\cite{OCRuCl3}.
The experimental fact that the intensity of the low-energy peak near 0.3 eV is 
much weaker than that of the dominant peak around 1.2 eV suggests 
that the strength of $t'$ is not as large as the theoretical estimation~\cite{Supp}.

{\it Conclusion  ---} 
Based on the numerically exact diagonalization analyses of the three-band Hubbard model, 
we have shown that 
the ground state of the $t_{2g}^5$ system with the honeycomb lattice structure 
can be transferred from the quasimolecular band insulator to the relativistic Mott insulator 
with increasing $\lambda$ as well as $U$ 
when the QMOs 
are disturbed and eventually replaced by the local relativistic $j_{\rm eff}$ orbitals.
We have demonstrated that the different electronic nature of these insulators is 
manifested in the electronic excitations observed in OC and RIXS. 
Comparing our results with experiments, we predict that 
not only Na$_2$IrO$_3$ with strong SOC but also $\alpha$-RuCl$_3$ with moderate SOC 
is the relativistic Mott insulator. 
This deserves further experimental confirmation, especially for $\alpha$-RuCl$_3$ where 
we expect the exciton-like excitation near the edge of the local $d$-$d$ excitation 
in the RIXS spectrum.

{\it Acknowledgements ---}
The numerical computations have been performed with 
the RIKEN supercomputer system (HOKUSAI GreatWave).
This work has been supported 
by Grant-in-Aid for Scientific Research from MEXT Japan 
under the Grant No. 25287096 and also 
by RIKEN iTHES Project and Molecular Systems.


\renewcommand{\thetable}{S\arabic{table}} 
\renewcommand{\thefigure}{S\arabic{figure}}
\renewcommand{\thetable}{S\arabic{table}}
\renewcommand\theequation{S\arabic{equation}}
\setcounter{table}{0}
\setcounter{figure}{0}
\setcounter{equation}{0}

\onecolumngrid

\clearpage

\begin{center}
{\bf \Large
\textit{Supplementary Materials}\\
From a quasimolecular band insulator to a relativistic Mott insulator 
in $t_{2g}^5$ systems with a honeycomb lattice structure}

\vspace{0.2 cm}

{\large
Beom Hyun Kim,$^{1,2}$ Tomonori Shirakawa,$^{3}$ and Seiji Yunoki$^{1,2,3,4}$
}

\vspace{0.1 cm}

{\it
$^1$Computational Condensed Matter Physics Laboratory, 
RIKEN, Wako, Saitama 351-0198, Japan

$^2$Interdisciplinary Theoretical Science (iTHES) Research Group,
RIKEN, Wako, Saitama 351-0198, Japan

$^3$Computational Quantum Matter Research Team, 
RIKEN Center for Emergent Matter Science (CEMS), Wako, Saitama 351-0198, Japan

$^4$Computational Materials Science Research Team, 
RIKEN Advanced Institute for Computational Science (AICS), 
Kobe, Hyogo 650-0047, Japan
}
\end{center}

\section{Hamiltonian}

\begin{figure}[b]
\centering
\includegraphics[width=.8\columnwidth]{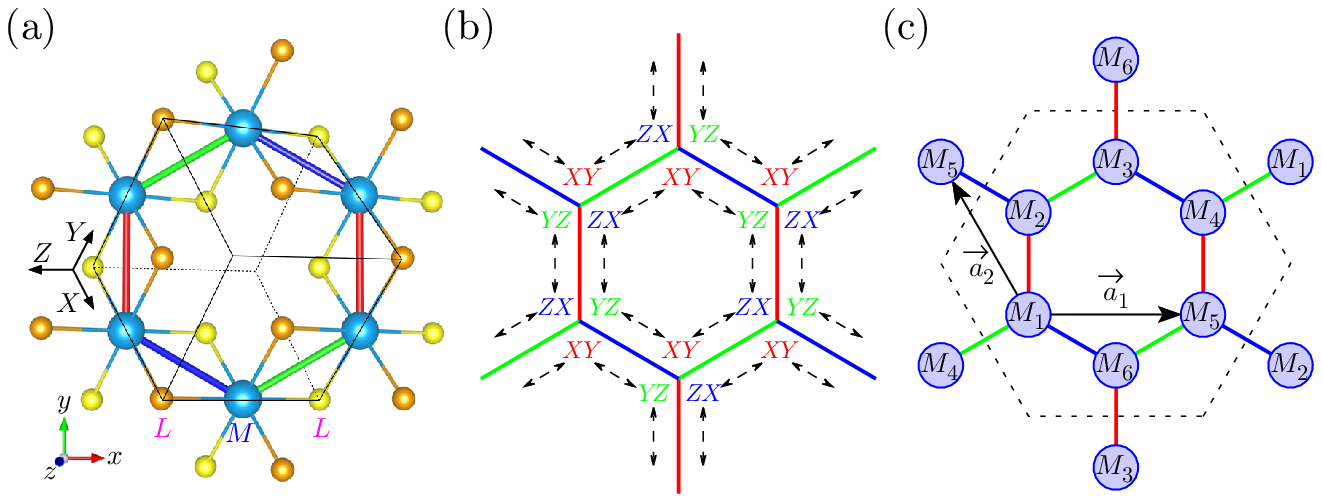}
\caption
{ 
  (a) Schematic crystal structure of the honeycomb lattice TM systems, 
  (b) its accessible hopping between neighboring TMs mediated via two edge-sharing ligands, and 
  (c) a periodic six-site cluster 
  (containing $M_1,M_2,\dots$, and $M_6$ sites surrounded by dashed lines) employed for the calculations. 
  TMs ($M$) are denoted by blue spheres and ligands ($L$) above (below) 
  the hexagon plane formed by TMs are indicated 
  by orange (yellow) spheres in (a). 
  $XY$, $YZ$, and $ZX$ in (b) refer to three $t_{2g}$ orbitals 
  ($d_{XY}$, $d_{YZ}$, and $d_{ZX}$, respectively) in the local 
  coordinates $X$, $Y$, and $Z$ defined in (a). 
  As indicated by double head arrows in (b), the electron motion is bounded in each hexagon 
  due to the bond-dependent hoppings when the SOC and Coulomb interaction are both absent.
  Red, green, and blue bonds in (a)--(c) represent the $XY$-, $YZ$-, and $ZX$-directions,
  respectively (also see TABLE~\ref{hop}). 
  The unit lattice vectors $\vec{a}_1=\sqrt{3}a\hat{x}$ and
  $\vec{a}_2=-\frac{\sqrt{3}a}{2}\hat{x}+\frac{3a}{2}\hat{y}$ 
  of the honeycomb lattice are also indicated in (c), where the global coordinates $x$, $y$, and $z$ are 
  given in (a), and $\hat x$, $\hat y$, and $\hat z$ are unit vectors along the $x$-, $y$-, and $z$-directions, 
  respectively.
}
\label{sfig1}
\end{figure}

As shown in Fig.~\ref{sfig1}(a), transition metal (TM) ions in 
the honeycomb lattice systems studied in the main text such as 
Na$_2$IrO$_3$ and $\alpha$-RuCl$_3$ 
are connected via two edge-sharing ligands.
To describe the electronic structure of these systems,
we adopt the three-band Hubbard Hamiltonian in the honeycomb lattice defined as 
\begin{align}
H =& \sum_{i, \delta, \mu, \nu, \sigma} 
 t_{\mu \nu}^{\delta}
 c_{i_\delta \mu \sigma}^{\dagger} c_{i \nu \sigma} +
   \lambda \sum_{i,\mu,\nu,\sigma,\sigma^{\prime}}
   (\mathbf{l}\cdot\mathbf{s})_{\mu\sigma,\nu\sigma^{\prime}} 
   c_{i\mu\sigma}^{\dagger}c_{i\nu\sigma^{\prime}}  \nonumber \\
  & + \frac{1}{2}\sum_{i,\sigma,\sigma^{\prime},\mu,\nu} 
  U_{\mu\nu} c_{i\mu\sigma}^{\dagger}
  c_{i\nu\sigma^{\prime}}^{\dagger}
  c_{i\nu\sigma^{\prime}}c_{i\mu\sigma}
  + \frac{1}{2}\sum_{\substack{i, \sigma, \sigma^{\prime},\mu,\nu \\ (\mu\ne\nu)}}
  J_{\mu\nu}c_{i\mu\sigma}^{\dagger}
  c_{i\nu\sigma^{\prime}}^{\dagger}
  c_{i\mu\sigma^{\prime}}c_{i\nu\sigma}
  + \frac{1}{2}\sum_{\substack{i, \sigma,\mu,\nu \\ (\mu\ne\nu)}} 
  J_{\mu\nu}^{\prime} c_{i\mu\sigma}^{\dagger}
  c_{i\mu\bar{\sigma}}^{\dagger}
  c_{i\nu\bar{\sigma}} c_{i\nu\sigma},  
\label{CLH}
\end{align}
where $c_{i\nu\sigma}^{\dagger}$ is the creation operator of an electron
with orbital $\nu\, (=XY,YZ,ZX)$ and spin $\sigma\,(=\uparrow,\downarrow)$ at site $i$ in the honeycomb lattice, 
and three $t_{2g}$ orbitals $XY$, $YZ$, and $ZX$ correspond respectively to $d_{XY}$, $d_{YZ}$, and $d_{ZX}$ 
orbitals defined in the local octahedron coordinates ($X$, $Y$, and $Z$),
as indicated in Fig.~\ref{sfig1}(a).
$\delta$ in the first term denotes the three nearest neighbor (NN) directions in the honeycomb lattice 
(i.e., $XY$-, $YZ$-, and $ZX$-directions, as indicated in Fig.~\ref{sfig1})
and $i_\delta$ is the $\delta$-directional NN site from site $i$. 
$\bar{\sigma}$ stands for the opposite spin of $\sigma$. 
The NN hopping matrix $t_{\mu \nu}^{\delta}$ is given in TABLE~\ref{hop} 
and the nonzero NN hoppings are schematically summarized in Fig.~\ref{sfig1}(b).
The second term in the right hand side of Eq.~(\ref{CLH}) describes the spin-orbit coupling (SOC) with 
the coupling constant $\lambda$, where $\mathbf l$ and $\mathbf s$ denote the orbital and spin angular 
momentum operators, respectively. 
The last three terms in the right hand side of Eq.~(\ref{CLH}) represent the two-body Coulomb interactions, 
where we assume the Kanamori-type Coulomb interaction, i.e., 
$U_{\mu\mu}=U$, $U_{\mu\ne\nu}=U-2J_{\rm H}$, and 
$J_{\mu\nu}=J_{\mu\nu}^{\prime}=J_{\rm H}$~\cite{Kanamori63S}.

\begin{table}[tb]
\caption
{
Hopping matrix $t_{\mu\nu}^\delta$ along three different directions.
The $XY$-, $YZ$-, and $ZX$-directions refer to the bond directions 
along which the NN TMs 
are connected via the two edge-sharing ligands located in the 
$XY$-, $YZ$-, and $ZX$-planes, respectively [see Fig.~\ref{sfig1}(a)]. 
$|ZX\rangle$, $|XY\rangle$, and $|YZ\rangle$ represent the three $t_{2g}$ orbitals, i.e.,  
$d_{ZX}$, $d_{XY}$, and $d_{YZ}$, respectively, defined in the local coordinates 
$X$, $Y$, and $Z$, as indicated in Fig.~\ref{sfig1}(a).
}
\label{hop}
\begin{ruledtabular}
\begin{tabular}{ c c c c c }
 & $XY$-direction & $YZ$-direction & $ZX$-direction & \\
 \hline
 &
 $\begin{matrix}
   & \vert ZX \rangle & \vert XY \rangle & \vert YZ \rangle\\
 \langle ZX \vert  &  0  & 0  & t \\
 \langle XY \vert  &  0  & 0  & 0 \\
 \langle YZ \vert  &  t  & 0  & 0 
 \end{matrix}$ &
 $\begin{matrix} 
   & \vert ZX \rangle & \vert XY \rangle & \vert YZ \rangle\\
 \langle ZX \vert  &  0  & t  & 0 \\
 \langle XY \vert  &  t  & 0  & 0 \\
 \langle YZ \vert  &  0  & 0  & 0 
 \end{matrix}$ &
 $\begin{matrix} 
   & \vert ZX \rangle & \vert XY \rangle & \vert YZ \rangle\\
 \langle ZX \vert  &  0  & 0  & 0 \\
 \langle XY \vert  &  0  & 0  & t \\
 \langle YZ \vert  &  0  & t  & 0 
 \end{matrix}$ &
\end{tabular}
\end{ruledtabular}
\end{table}

Let us define the local orbitals $\tilde{y}$, $\tilde{z}$, and $\tilde{x}$ as 
\begin{eqnarray}
|\tilde{y}\rangle &=& \frac{1}{\sqrt{2}} \left( |ZX\rangle - |YZ\rangle \right), \label{Oy} \\
|\tilde{z}\rangle &=& \frac{1}{\sqrt{3}} \left( |ZX\rangle + |XY\rangle + |YZ\rangle \right),  \label{Oz} \\
|\tilde{x}\rangle &=& \frac{1}{\sqrt{6}} \left( |ZX\rangle -2|XY\rangle + |YZ\rangle \right) \label{Ox} ,
\end{eqnarray}
which correspond to $\sqrt{\frac{2}{3}} d_{xy}\!-\!\sqrt{\frac{1}{3}} d_{yz}$, $d_{z^2}$,
and $-\sqrt{\frac{2}{3}} d_{x^2-y^2}\!-\!\sqrt{\frac{1}{3}} d_{zx}$, respectively,
in the global coordinates $x$, $y$, and $z$ of the honeycomb lattice [see Fig.~\ref{sfig1}(a)]. 

Notice that the unitary transformation given above from $ZX$, $XY$, and $YZ$ orbitals to
$\tilde{y}$, $\tilde{z}$, and $\tilde{x}$ orbitals
is exactly the same as that for $p$ orbitals 
from $p_Y$, $p_Z$, and $p_X$ orbitals in the local coordinates to 
$p_y$, $p_z$, and $p_x$ orbitals in the global coordinates.
With these orbitals as the basis set, the SOC is nonzero 
only for
$\langle \tilde{x} \sigma|\mathbf{l}\cdot\mathbf{s}|\tilde{y}\sigma \rangle
= \frac{i}{2}s_\sigma$,
$\langle \tilde{y}\sigma|\mathbf{l}\cdot\mathbf{s}|\tilde{z}\bar{\sigma}\rangle
= \frac{i}{2}$, and
$\langle \tilde{z}\sigma|\mathbf{l}\cdot\mathbf{s}|\tilde{x}\bar{\sigma}\rangle
=\frac{1}{2}s_\sigma$, where $s_\sigma=1\,(-1)$ for $\sigma=\uparrow\, (\downarrow)$ and 
the imaginary unit $i$ should not be confused with the site index $i$ in Eq.~(\ref{CLH}). 
The local SOC term can be readily diagonalized in terms of the relativistic orbitals with the effective 
total angular momentum $j_{\rm eff}=1/2$, i.e.,  
\begin{align}
\vert j_{\rm eff}=1/2,j_{\rm eff}^z=+1/2 \rangle &=- \frac{1}{\sqrt{3}} \left( 
 \vert \tilde{z}+ \rangle + \vert  \tilde{x}- \rangle  
 + i \vert  \tilde{y} -\rangle  \right), \label{jph}\\
\vert j_{\rm eff}=1/2,j_{\rm eff}^z=-1/2 \rangle &=  +\frac{1}{\sqrt{3}} \left( 
 \vert \tilde{z}- \rangle - \vert  \tilde{x} +\rangle 
 +i \vert  \tilde{y}+\rangle  \right), \label{jmh}
\end{align}
and those with $j_{\rm eff}=3/2$, i.e., 
\begin{align}
\vert j_{\rm eff}=3/2,j_{\rm eff}^z=+3/2 \rangle &= -\frac{1}{\sqrt{2}} \left( 
\vert  \tilde{x}+\rangle  + i \vert  \tilde{y} +\rangle  \right),\\
\vert j_{\rm eff}=3/2,j_{\rm eff}^z=+1/2 \rangle &=+\frac{1}{\sqrt{6}} \left( 
 2 \vert \tilde{z}+ \rangle -\vert  \tilde{x}- \rangle  
- i \vert  \tilde{y} -\rangle  \right), \\
\vert j_{\rm eff}=3/2,j_{\rm eff}^z=-1/2 \rangle &=+ \frac{1}{\sqrt{6}} \left( 
2  \vert \tilde{z}- \rangle +\vert  \tilde{x} +\rangle 
-i \vert  \tilde{y}+\rangle  \right),\\
\vert j_{\rm eff}=3/2,j_{\rm eff}^z=-3/2 \rangle &=+\frac{1}{\sqrt{2}} \left( 
\vert  \tilde{x}-\rangle  - i \vert  \tilde{y} -\rangle  \right),
\end{align}
whose energies are $\lambda$ for $j_{\rm eff}=1/2$ and 
$-\lambda/2$ for $j_{\rm eff}=3/2$.

\section{Quasimolecular bands}

Let us define the unit lattice vectors $\vec{a}_1=\sqrt{3}a\hat{x}$ and
$\vec{a}_2=-\frac{\sqrt{3}a}{2}\hat{x}+\frac{3a}{2}\hat{y}$ of the honeycomb lattice [see Fig.~\ref{sfig1}(c)], 
where $a$ is the lattice constant between NN sites. Note that the unit cell contains two basis sites, 
``1" and ``2", located at $\mathbf{0}$ and $a\hat{y}$, respectively. 
The unit reciprocal lattice vectors are thus given as 
$\vec{b}_1=\frac{4\pi}{2\sqrt{3}a}
  \big( \hat{x} +\frac{1}{\sqrt{3}} \hat{y} \big)$
and $\vec{b}_2=\frac{4\pi}{3a}\hat{y}$.

In the non-interacting limit with $\lambda=0$, the Hamiltonian $H$ can be expressed simply by the 
following tight-binding Hamiltonian 
\begin{equation}
H_t = \sum_{\mathbf k} \sum_{\alpha,\beta}\sum_{i,j} \sum_{ \sigma } 
 H_{\alpha_i \beta_j} (\mathbf{k}) c_{\alpha_i\sigma}^{\dagger} (\mathbf{k}) 
 c_{\beta_j\sigma}(\mathbf{k}),
\label{TBr}
\end{equation}
where $c_{\alpha_i\sigma}^{\dagger} (\mathbf{k})$ is the Fourier transformation
of the creation operator of an electron with 
orbital $\alpha$ and spin $\sigma$ at the $i$-th basis site ($i=1,2$). 
The Hamiltonian matrix $H(\mathbf{k})$ is given as 
\begin{equation}
H(\mathbf{k}) =
\begin{pmatrix}
 0 & 0 & 0 & 0 & t e^{ i (k_1+k_2)}& t \\
 0 & 0 & 0 & t e^{ i (k_1+k_2)} & 0 & t e^{i k_2 } \\
 0 & 0 & 0 & t &t e^{i k_2}  & 0 \\
 0 & t e^{- i (k_1+k_2)} & t & 0 & 0 & 0 \\
  t e^{- i (k_1+k_2)} & 0 &   t e^{-i k_2} & 0 & 0 & 0  \\
 t & t e^{-i k_2} & 0 & 0 & 0 & 0  
\end{pmatrix} \label{Hk}
\end{equation}
in the basis of $( ZX_1,XY_1,YZ_1,ZX_2,XY_2,YZ_2 )$ for wave number 
$\mathbf{k}=\frac{k_1}{2\pi} \vec{b}_1+ \frac{k_2}{2\pi}\vec{b}_2$ with $k_1$ and 
$k_2$ being real numbers.

Introducing the $(3\times 3)$ matrix 
\begin{equation}
A(\mathbf{k})=
\begin{pmatrix}
0 & t e^{ i (k_1+k_2)}& t \\
 t e^{ i (k_1+k_2)} & 0 & t e^{i k_2 } \\
 t &t e^{i k_2}  & 0 
\end{pmatrix},
\end{equation}
the Hamiltonian matrix $H(\mathbf{k})$ is expressed as
\begin{equation}
H(\mathbf{k}) =
\begin{pmatrix}
 0 & A(\mathbf{k}) \\
 A(\mathbf{k})^{\dagger} & 0 
\end{pmatrix}.
\end{equation}
Let the eigenvalue and the eigenstate of $H(\mathbf{k})$ be $E$ and 
$\mathbf{X}_E = \begin{pmatrix} X & Y \end{pmatrix}^{\textrm{T}}$, respectively.
Because 
\begin{equation}
\begin{pmatrix}
 0 & A(\mathbf{k}) \\
 A(\mathbf{k})^{\dagger} & 0 
\end{pmatrix}
\begin{pmatrix}
 X \\
 Y
\end{pmatrix} 
=E
\begin{pmatrix}
 X \\
 Y
\end{pmatrix} 
,
\end{equation}
$A(\mathbf{k}) Y = E X$ and $A(\mathbf{k})^{\dagger} X = E Y$.
We can thus rewrite the eigenvalue problem as 
\begin{equation}
A(\mathbf{k}) A(\mathbf{k})^{\dagger} X = t^2
\begin{pmatrix}
  2 & e^{- i k_2} &  e^{ i k_1} \\
  e^{  i k_2}  & 2 & e^{i(k_1+k_2 )} \\
  e^{- i k_1} & e^{-i(k_1+k_2)}  & 2 
\end{pmatrix} X
= E^2 X.
\label{AE}
\end{equation}
By solving the above equation, we can obtain the following eigenstates
\begin{align}
& X_1^{\textrm{T}}= \begin{pmatrix}
1, & e^{i k_2}, & e^{-i k_1}
\end{pmatrix}, \\
& X_2^{\textrm{T}}=
\begin{pmatrix}
0, & e^{i k_2}, & - e^{-i k_1}
\end{pmatrix}, \\
& X_3^{\textrm{T}}=
\begin{pmatrix}
-2, & e^{i k_2}, & e^{-i k_1}
\end{pmatrix},
\end{align}
whose eigenvalues $E^2$ are $4t^2$, $t^2$, and $t^2$, respectively.
Since $Y=\frac{1}{E}A(\mathbf{k})^{\dagger} X$,
we finally obtain the six eigenstates of $H(\mathbf{k})$ as 
\begin{align}
\label{a1g}
&\mathbf{X}_{a_{1g}}(\mathbf{k}) = \frac {1}{\sqrt{6}} 
\begin{pmatrix}
1, & e^{ i k_2 }, & e^{-i k_1 }, &
 e^{- i k_1 }, & e^{- i (k_1+k_2) }, & 1 \end{pmatrix}^{\textrm{T}}, \\
&\mathbf{X}_{e_{2u,1}}(\mathbf{k}) = \frac {1}{2}  
\begin{pmatrix}
 0, & e^{ i k_2 }, & -e^{-i k_1}, & 
 0, & - e^{- i (k_1+k_2)}, & 1 \end{pmatrix}^{\textrm{T}}, \\
&\mathbf{X}_{e_{2u,2}}(\mathbf{k}) = \frac {1}{2\sqrt{3}} 
\begin{pmatrix} 
 -2, & e^{i k_2} ,& e^{-i k_1}, & 
  2e^{-i k_1}, & -e^{-i(k_1+k_2)}, & -1  \end{pmatrix}^{\textrm{T}}, \\
&\mathbf{X}_{e_{1g,1}}(\mathbf{k}) = \frac {1}{2 }  
\begin{pmatrix} 
 0, & e^{i k_2}, & -e^{-i k_1}, &
 0, & e^{-i(k_1+k_2)}, & -1 \end{pmatrix}^{\textrm{T}},  \\
&\mathbf{X}_{e_{1g,2}}(\mathbf{k}) = \frac {1}{2\sqrt{3}} 
\begin{pmatrix} 
 -2, & e^{i k_2}, & e^{-i k_1}, & 
 -2e^{-i k_1}, & e^{-i(k_1+k_2)}, & 1  \end{pmatrix}^{\textrm{T}},  \\
\label{b1u}
&\mathbf{X}_{b_{1u}}(\mathbf{k}) = \frac {1}{\sqrt{6}} 
\begin{pmatrix}
 1, & e^{ i k_2 }, & e^{-i k_1 }, &
-e^{- i k_1 }, & -e^{-i(k_1+k_2) }, & -1 \end{pmatrix}^{\textrm{T}},  
\end{align}
with the eigenvalues being $2t$, $t$, $t$, $-t$, $-t$, and $-2t$, respectively.
Because all eigenvalues are wave number independent, we can call these energy bands 
as quasimolecular bands formed by quasimolecular orbitals (QMOs) with their symmetries, 
$a_{1g}$, $e_{2u}$, $e_{1g}$, and $b_{1u}$, 
indicated in the above eigenstates.

\section{cluster calculations}

To explore the correlation effect, we consider a periodic six-site cluster shown 
in Fig~\ref{sfig1}(c), which contains six holes for the $t_{2g}^5$ systems discussed in the main text. 
We take into account all states to represent the whole Hilbert space and thus 
the size of the Hilbert space is $_{36}C_{6}=1947792$.
To solve Eq.~(\ref{CLH}) and find the ground state, we employ 
the exact diagonalization method based on Lanczos algorithm 
with the Jacobi preconditioner~\cite{Morgan93S}. 
We obtain the ground state and the corresponding eigenenergy
with the energy accuracy of $4\times10^{-10}t$.

\subsection{Hole density of $a_{1g}$ band}

Since only the $a_{1g}$ band is fully empty in the non-relativistic and non-interacting limits for 
the $t_{2g}^5$ systems, the hole density in the $a_{1g}$ quasimolecular band 
is a good quantity to quantify how far the ground state is from the pure QMO state. 
The hole density operator $\bar{\rho}_{a_{1g}} (\mathbf{k})$ per site at the wave number $\mathbf k$ is defined as
\begin{equation}
\bar{\rho}_{a_{1g}} (\mathbf{k}) = 
\frac{1}{2}\sum_{\sigma} 
c_{a_{1g}\sigma}(\mathbf{k})c^{\dagger}_{a_{1g}\sigma}(\mathbf{k}),
\end{equation}
where the creation operator $c^{\dagger}_{a_{1g}\sigma}(\mathbf{k})$ is the eigenstate of $H({\mathbf k})$ 
and is given as 
$
c^{\dagger}_{a_{1g}\sigma}(\mathbf{k}) = 
\sum_{j,\nu} [\mathbf{X}_{a_{1g}}(\mathbf{k})]_{\nu_j} 
c^{\dagger}_{\nu_j\sigma}(\mathbf{k}).
$
Here $[\mathbf{X}_{a_{1g}}(\mathbf{k})]_{\nu_j}$ ($j=1,2$ and $\nu=ZX, XY,YZ$) is 
the $\nu_j$-th element of the eigenstate of $H(\mathbf{k})$ with $a_{1g}$ symmetry given 
in Eq.~(\ref{a1g}).
For the six-site cluster with periodic boundary conditions, 
there exist three independent wave numbers, i.e., 
$(k_1,k_2)=(0,0)$ ($\Gamma$ point), 
$(\frac{\pi}{3},\frac{\pi}{3})$ ($K$ point), 
and $(\frac{2\pi}{3},\frac{2\pi}{3})$ ($K'$ point), and 
$c^{\dagger}_{a_{1g}\sigma}(\mathbf{k})$ for these wave numbers are 
expressed as 
\begin{align}
c^{\dagger}_{a_{1g}\sigma}(\mathbf{k}) &= \frac {1}{\sqrt{3}}  
\Big(
\sum_{\nu} [\mathbf{X}_{a_{1g}}(\mathbf{k})]_{\nu_1} 
  c^{\dagger}_{1\nu\sigma} +
\sum_{\nu} [\mathbf{X}_{a_{1g}}(\mathbf{k})]_{\nu_2}
  c^{\dagger}_{2\nu\sigma} +
\sum_{\nu} [\mathbf{X}_{a_{1g}}(\mathbf{k})]_{\nu_1}
  e^{-i(k_1+k_2)} c^{\dagger}_{3\nu\sigma} \nonumber \\
&+
\sum_{\nu} [\mathbf{X}_{a_{1g}} (\mathbf{k})]_{\nu_2} 
 e^{-ik_1} c^{\dagger}_{4\nu\sigma} +
\sum_{\nu} [\mathbf{X}_{a_{1g}}(\mathbf{k})]_{\nu_1}
 e^{-ik_1} c^{\dagger}_{5\nu\sigma} +
\sum_{\nu} [\mathbf{X}_{a_{1g}}(\mathbf{k})]_{\nu_2}
 e^{ ik_2} c^{\dagger}_{6\nu\sigma}
\Big),
\end{align}
where $c^{\dagger}_{j\nu\sigma}$ is the creation operator of an electron with orbital 
$\nu=(YZ,ZX,XY)$ and spin $\sigma=(\uparrow,\downarrow)$ at site $M_j$ in the six-site
cluster shown in Fig.~\ref{sfig1}(c). 
In the main text, we calculate the hole density at the $\Gamma$ point, i.e., 
\begin{equation}
\bar{n}_{a_{1g}}=
\langle \Psi_0 \vert \bar{\rho}_{a_{1g}}(\mathbf{k}=0) 
\vert \Psi_0 \rangle,
\end{equation} 
where $\vert \Psi_0 \rangle$ is the ground state of the six-site cluster.
This quantity $\bar{n}_{a_{1g}}$ is exactly one 
for the pure QMO state.

\subsection{Doublon-holon excitations in $j_{\rm eff}=1/2$ orbitals}

To explore the distribution of a doublon-holon pair in the 
$j_{\rm eff}=1/2$ orbitals at neighboring sites,
we calculate the projected spectrum 
\begin{align}
\Lambda_{1/2} (\omega) &=  \sum_n \sum_{s_3,s_4,s_5,s_6} 
\left | \langle D_{s_3,s_4,s_5,s_6} | \Psi_n\rangle \right |^2 \delta(\omega-E_n+E_0) \nonumber \\
 &= - \frac{1}{\pi} \textrm{Im} \sum_{s_3,s_4,s_5,s_6} 
\langle D_{s_3,s_4,s_5,s_6} | \frac{1}{\omega-H+E_0+i\delta}| D_{s_3,s_4,s_5,s_6} \rangle,
\end{align}
where $|\Psi_n\rangle$ and $E_n$ are the $n$-th eigenstate and eigenvalue of $H$, respectively, and 
the ground state corresponds to $n=0$. 
$| D_{s_3,s_4,s_5,s_6} \rangle$ is a single doublon-holon pair excited state in the NN $j_{\rm eff}=1/2$ orbitals 
at sites $M_1$ and $M_2$ [see Fig.~\ref{sfig1}(c)], defined as 
\begin{equation}
| D_{s_3,s_4,s_5,s_6} \rangle = 
d_{1,+1/2}d_{1,-1/2} d_{3,s_3}d_{4,s_4}
d_{5,s_3}d_{6,s_6} \vert  \textrm{FO} \rangle,
\end{equation}
where $\vert  \textrm{FO} \rangle$ is the fully occupied state with electrons and 
$d_{l,s_l}$ is the annihilation operator of $j_{\rm eff}=1/2$ electron with isospin $s_l\,(=\pm1/2)$ 
[given in Eqs.~(\ref{jph}) and (\ref{jmh})] at site $M_l$ in the six-site cluster shown in Fig.~\ref{sfig1}(c). 
To calcuate $\Lambda_{1/2}(\omega)$, 
we employ the continued faction method~\cite{Dagotto94S}.
We perform 500 iterations of the continued faction with 
$\delta=0.12t$.

\begin{figure}[h]
\centering
\includegraphics[width=.65\columnwidth]{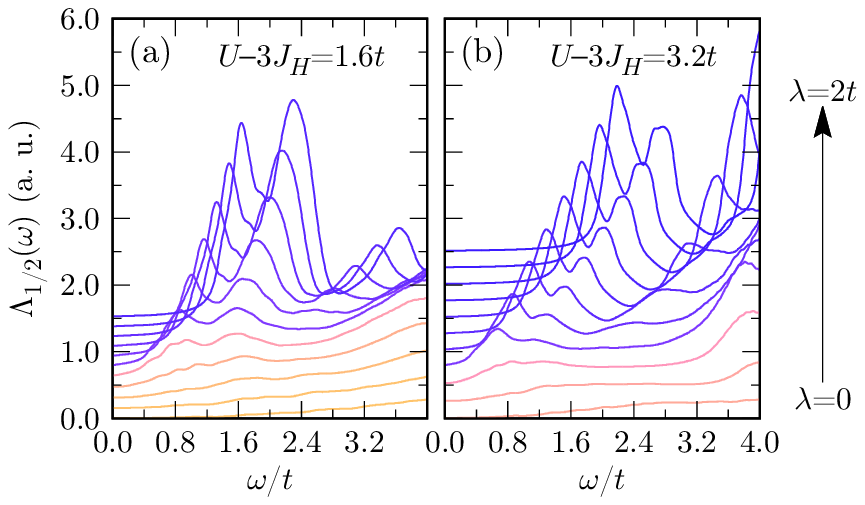}
\caption
{ (Color online)
Doublon-holon excitation spectrum $\Lambda_{1/2} (\omega)$ for 
$U-3J_{\rm H}= 1.6t$ (a) and $3.2t$ (b) 
with varying $\lambda$ indicated in the figure. 
Here we set $J_{\rm H}=1.6t$ and different 
line colors correspond to the values of 
$\bar n_{a_{1g}}$ shown in Fig.~2(b) of the main text. 
}
\label{lambda2}
\end{figure}

As shown in Figs.~3(a)--(c) of the main text, the $U$ dependence of the low-energy peak positions 
in $\Lambda_{1/2} (\omega)$ for given $\lambda$ 
is qualitatively different in the two insulators, i.e., the quasimolecular band insulator and 
the relativistic Mott insulator. 
As shown in Fig.~\ref{lambda2}, the similar change of 
behavior is found even when $\lambda$ is varied for fixed $U$ and $J_{\rm H}$. 
With increasing $\lambda$, 
the low-energy peak positions first shift downward in energy and then move upward once we cross 
the crossover region. 
The increase of the low-energy peak positions supports that the edge of
$\Lambda_{1/2}(\omega)$ in the relativistic Mott limit is evidently coupled with the local $d$-$d$
excitation, whose excitation energy is proportional to $\lambda$.
Therefore, the exciton-like peak as well as the spin-orbit exciton peak in the RIXS spectrum 
should also shift upward with increasing $\lambda$.
The decrease of the 
low-energy peak positions in $\Lambda_{1/2}(\omega)$ in the QMO limit is due to the decrease of a band gap 
with increasing $\lambda$, as shown in Fig.~1(b) of the main text.

\subsection{Optical conductivity}

We adopt the Kubo formula to calculate the optical conductivity (OC), 
\begin{align}
\sigma(\omega) &= \frac{\pi v}{\omega}
  \sum_{n (\ne 0)} |\langle \Psi_n|\hat J_c|\Psi_0\rangle|^2 
  \delta (\omega-E_n\!+\!E_0) \nonumber \\
   &= - v \ \textrm{Im} 
  \sum_{n (\ne 0)} \frac{|\langle \Psi_n|\hat J_c|\Psi_0\rangle|^2} 
  {(E_n-E_0)(\omega-E_n\!+\!E_0+i\delta)} \nonumber \\
   &= - v \ \textrm{Im} \Big[ \frac{1}{\omega+i\delta}
  \sum_{n (\ne 0)} \Big( \frac{|\langle \Psi_n|\hat J_c|\Psi_0\rangle|^2} 
  {E_n-E_0} + 
   \frac{|\langle \Psi_n|\hat J_c|\Psi_0\rangle|^2}
   {\omega-E_n\!+\!E_0+i\delta}\Big) \Big] \nonumber \\
   &= - v \ \textrm{Im} \Big[ \frac{1}{\omega+i\delta}
   \Big(\langle \Psi_0| \hat J_c \frac{1}{H-E_0} \hat J_c|\Psi_0\rangle
  + \langle \Psi_0| \hat J_c \frac{1}{\omega-H+E_0+i\delta} 
  \hat J_c|\Psi_0\rangle \Big) \Big],
\label{eqOC}
\end{align}
where $v$ is the volume per site and $\hat J_c$ 
is the current operator along the $XY$-direction given as 
\begin{equation}
\hat J_c = -\frac{i e}{A\hbar} \sum_{\mu,\nu,\sigma} 
\big( t^{XY}_{\mu\nu}
 c_{2\nu\sigma}^{\dagger} c_{1\mu\sigma}
- t^{XY}_{\mu\nu}
 c_{1\mu\sigma}^{\dagger} c_{2\nu\sigma}
\big),
\end{equation}
where $e\, (<0)$ is the electronic charge and $A$ is the area facing
neighboring sites. 
Note that the OCs for the $YZ$- and $ZX$-directions are exactly the same as that for the $XY$-direction 
because of the symmetry. 
We employ the continued fraction method to calculate the spectrum in Eq.~(\ref{eqOC}).
We perform $400$ iterations of the continued fraction with 
$\delta=0.4t$.

\subsection{Resonant inelastic x-ray scattering spectrum}

To calculate the resonant inelastic x-ray scattering (RIXS) spectrum $I\left(\omega,\mathbf{q},\bm{\epsilon},\bm{\epsilon}^{\prime}\right)$, 
we follow the Kramers-Heisenberg formula, i.e., 
\begin{align}
I\left(\omega,\mathbf{q},\bm{\epsilon},\bm{\epsilon}^{\prime}
  \right)=\sum_n |\mathcal{F}_{n0}(\omega,\mathbf{q},
 \bm{\epsilon},\bm{\epsilon}^{\prime})|^2
  \delta\left(\omega-E_n+E_0\right),
\end{align}
where $\omega=\omega_{\mathbf{k}}-\omega_{\mathbf{k}^{\prime}}$
and $\mathbf{q}=\mathbf{k}-\mathbf{k}^{\prime}$
when an incident (outgoing) x-ray has momentum $\mathbf{k}$ 
($\mathbf{k}^{\prime}$), energy $\omega_{\mathbf{k}}$ 
($\omega_{\mathbf{k}^{\prime}}$), 
and polarization $\bm{\epsilon}$ ($\bm{\epsilon}^{\prime}$)~\cite{Ament11S}.
Adopting the dipole and fast collision approximations,
$\mathcal{F}_{n0}(\omega,\mathbf{q},
 \bm{\epsilon},\bm{\epsilon}^{\prime})$ is described as
$
\mathcal{F}_{n0}(\omega,\mathbf{q},
 \bm{\epsilon},\bm{\epsilon}^{\prime})=\frac{1}{i\Gamma_b} \langle \Psi_n|
R(\bm{\epsilon}^{\prime},\bm{\epsilon},\mathbf{q}) | \Psi_0 \rangle
$, where $\Gamma_b$ is a lifetime broadening for the intermediate states. 
The RIXS scattering operator 
$R(\bm{\epsilon}^{\prime},\bm{\epsilon},\mathbf{q})$ is given as 
\begin{equation}
\label{R_eq}
R(\bm{\epsilon}^{\prime},\bm{\epsilon},\mathbf{q})=
\sum_i\sum_{\nu,\nu^{\prime},\sigma} e^{i\mathbf{q}\cdot\mathbf{r}_i}
 T_{\nu^{\prime}\nu}(\bm{\epsilon}^{\prime},\bm{\epsilon})
 c_{i\nu^{\prime}\sigma}c_{i\nu\sigma}^{\dagger},
\end{equation}
where 
   $T_{\nu^{\prime}\nu}(\bm{\epsilon}^{\prime},\bm{\epsilon})=
 \sum_{s} \langle \phi_s| \bm{\epsilon}^{\prime}\cdot \mathbf{r} 
 |\psi_{\nu^{\prime}} \rangle
  \langle \psi_{\nu} | \bm{\epsilon}\cdot \mathbf{r} |\phi_s\rangle$ and
$\mathbf{r}$ is the position operator of valence and core-hole electrons,
${\mathbf r}_i$ is the position vector of lattice site $i$,
$\psi_\nu$ is the local atomic wave function for $t_{2g}$ orbital $\nu$, and 
$\phi_s$ refers to the core-hole wave function ($2p_{3/2}$ for the $L_3$-edge spectrum).
The RIXS spectrum $I\left(\omega,\mathbf{q},\bm{\epsilon},\bm{\epsilon}^{\prime}\right)$ is finally given as 
\begin{equation}
I\left(\omega,\mathbf{q},\bm{\epsilon},\bm{\epsilon^{\prime}} \right)
 =-\frac{1}{\pi\Gamma_b} \textrm{Im}\Big[
 \langle \Psi_0| R(\bm{\epsilon}^{\prime},\bm{\epsilon},\mathbf{q})
 \frac{1}{\omega-H+E_0+i\delta}
 R(\bm{\epsilon}^{\prime},\bm{\epsilon},\mathbf{q}) |\Psi_0\rangle
\Big].
\end{equation}
For the calculations in the main text, 
we set $\bm{\epsilon}=\frac{1}{\sqrt{2}}\hat{x}+\frac{1}{\sqrt{2}}\hat{z}$,
$\bm{\epsilon}'=\hat{y}$, and
$\mathbf{q}=0$ where the incident and outgoing x-rays have 
$\pi$ and $\sigma$ polarizations, respectively, with 45$^{\circ}$ scattering angles
for the honeycomb plane.
The geometry is in principle the same as the experimental setup in Ref.~[\onlinecite{Chun2015S}].
We employ the continued fraction method to evaluate the RIXS spectrum and 
perform 300 iterations of the continued fraction with 
$\delta=0.08t$.

\begin{figure}[h]
\centering
\includegraphics[width=.8\columnwidth]{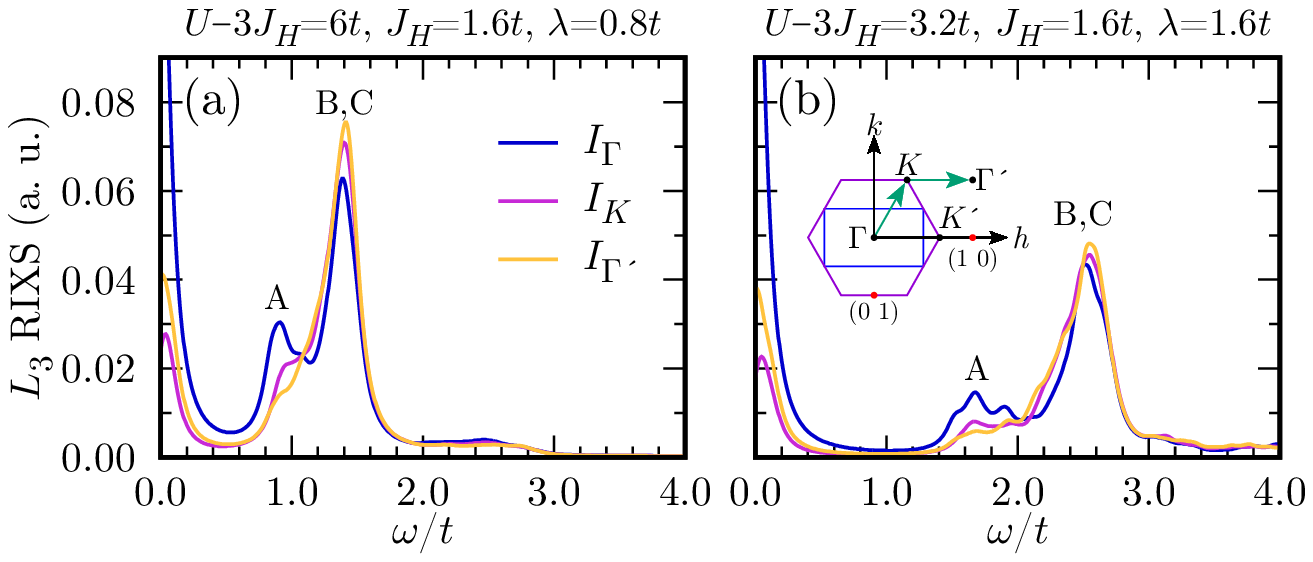}
\caption
{ (Color online) 
  The $L_3$-edge RIXS spectrum for (a) $U-3J_H=6t$ and 
  $\lambda=0.8t$, and (b) $U-3J_H=3.2t$ and $\lambda=1.6t$ at 
  three different momenta ($\Gamma$, $K$, and $\Gamma'$ points) in the Brillouin zone (BZ) of
  the honeycomb lattice indicated as the inset in (b). 
  We set $J_H=1.6t$. 
  Peaks labeled with `A', `B', and `C' correspond to those observed in the 
  $L_3$-edge RIXS experiment~\cite{Gretarsson2013S}.
  Because the trigonal distortion is not considered here, 
  peaks `B' and `C' tend to merge (see Fig.~\ref{rixs_dist}). 
  To facilitate the comparison with the experiment~\cite{Gretarsson2013S}, 
  the BZ for monoclinic Na$_2$IrO$_3$ is also indicated in the inset by a blue rectangle along with the 
  corresponding momentum coordinates $(h~k)$ 
  employed in Ref.~[\onlinecite{Gretarsson2013S}], where 
  two red points denote $(1~0)$ and $(0~1)$. 
  In the monoclinic coordinates, $\Gamma$, $K$, $K'$, and $\Gamma'$ can be expressed as
  $(0~0)$, $(\frac{1}{3}~1)$, $(\frac{2}{3}~0)$, and $(1~1)$, respectively.
  Note that, due to the $C_3$ rotation and inversion symmetries in our model, 
  the calculated RIXS spectrum at the $K'$ point is exactly same as that
  at the $K$ point.
}
\label{rixsq}
\end{figure}

We also examine the momentum dependence of the  $L_3$-edge RIXS spectrum 
with the same polarizations, i.e., $\pi$ and $\sigma$ polarizations for 
the incident and outgoing x-rays, respectively. 
Figure~\ref{rixsq} summarizes the results 
for three different momenta at the $\Gamma$, $K$, and $\Gamma'$ points 
when $U$ and $\lambda$ are both strong enough to stabilize 
the relativistic Mott insulator. 
Note that these three momenta are 
accessible irreducible momenta 
in the periodic six-site cluster employed with 
both $C_3$ rotation and inversion symmetry. 
According to the experimental observation in Ref.~[\onlinecite{Gretarsson2013S}],
the RIXS spectrum in Na$_2$IrO$_3$ shows three characteristic peaks 
(denoted as `A', `B', and `C' in Fig. 2 of Ref.~[\onlinecite{Gretarsson2013S}]) around the excitation
energy of the local $d$-$d$ transition from the $j_{\rm eff}=1/2$ to $3/2$ orbitals, where 
peaks `B' and `C' have dominant intensities and exhibit 
less dependence of momentum $\mathbf q$, whereas the intensity of peak `A' 
is almost diminished when the momentum is far from the $\Gamma$ point  
[here, do not confuse `A', `B', and `C' with labels used for the three types of excitations 
shown in Fig. 3(d) of the main text]. 
As discussed in the following (and also see Refs.~[\onlinecite{Gretarsson2013S}] and [\onlinecite{BHKim2014S}]), 
the splitting of peaks `B' and `C' are attributed to 
the trigonal distortion of the IrO$_6$ octahedron and these two peaks tend to merge 
in the absence of the trigonal distortion.
Our theoretical calculations shown in Fig.~\ref{rixsq} are performed without the trigonal distortion and 
therefore they are in good qualitative agreement with the experimental observation: 
while the dominant main peak (denoted as `B' and `C' in Fig.~\ref{rixsq}) with large intensity 
is almost momentum independent, the exciton-like peak (denoted as `A' in Fig.~\ref{rixsq}) is strongly 
suppressed at the $K$ and $\Gamma'$ points as compared with that at the $\Gamma$ point. 

\begin{figure}[h]
\centering
\includegraphics[width=.9\columnwidth]{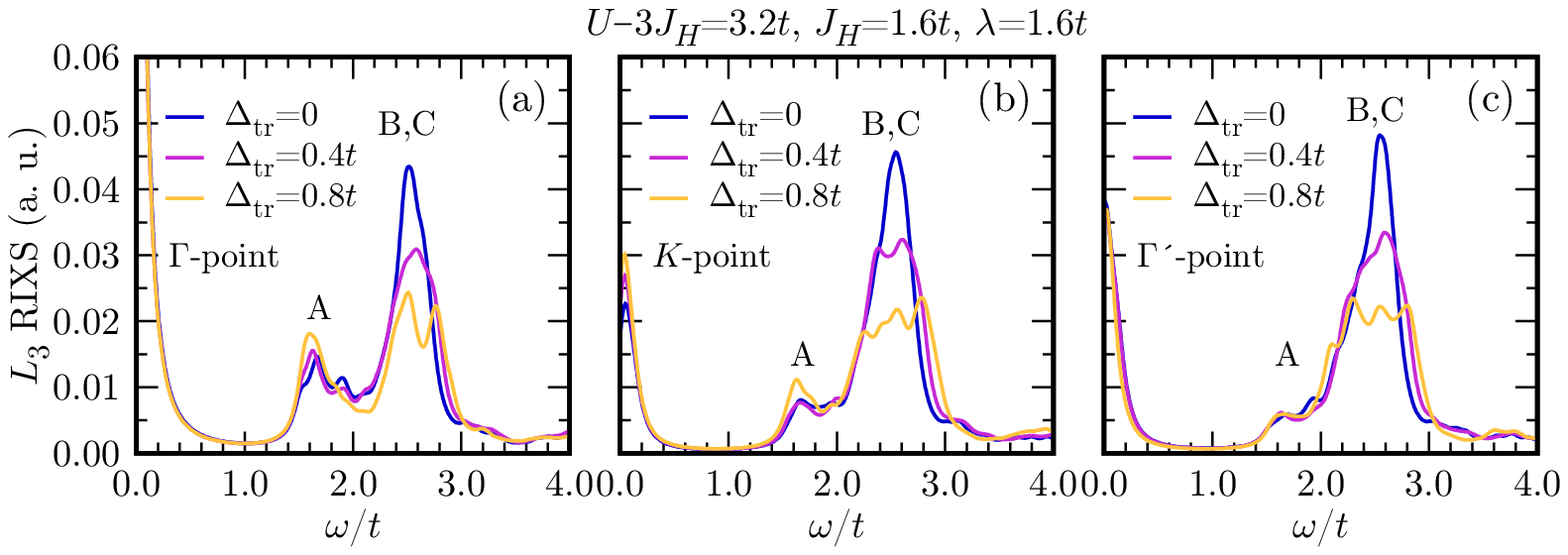}
\caption
{ (Color online)
  The $L_3$-edge RIXS spectrum for $U-3J_H=3.2t$, $J_H=1.6t$, and $\lambda=1.6t$
  calculated at (a) $\Gamma$, (b) $K$, and (c) $\Gamma'$ points 
  [see the inset of Fig.~\ref{rixsq}(b)] 
  for three different trigonal crystal fields 
  $\Delta_{\rm tr}$ denoted in the figure. 
  The main peak corresponding to the local $d$-$d$ transition is split into double (or multiple) peaks 
  (denoted as `B' and `C') in the presence of te trigonal distortion. 
}
\label{rixs_dist}
\end{figure}

Finally, we investigate the effect of the trigonal distortion on the RIXS spectrum. For this purpose, 
we introduce into the Hamiltonian $H$ in Eq.~(\ref{CLH}) the additional term $H_{\rm tr}$ representing the 
trigonal crystal field splitting among the $t_{2g}$ orbitals.
Since the local orbitals $\tilde x$, $\tilde y$, and $\tilde z$ introduced in Eq.~(\ref{Oy})--(\ref{Ox}) are 
already irreducible representations in the presence of the trigonal distortion with 
symmetries $a_{1g}$ ($\tilde{z}$) and $e_{g}'$ ($\tilde{x}$, $\tilde{y}$), 
the trigonal distortion is simply represented by the Hamiltonian 
\begin{equation}
H_{\rm tr} = \frac{\Delta_{\rm tr}}{3} \sum_{i\sigma}\Big( 
  c_{i\tilde{x}\sigma}^{\dagger} c_{i\tilde{x}\sigma}
 +c_{i\tilde{y}\sigma}^{\dagger} c_{i\tilde{y}\sigma}
 -2 c_{i\tilde{z}\sigma}^{\dagger} c_{i\tilde{z}\sigma}
\Big), 
\end{equation}
where the compressed distortion observed in Na$_2$IrO$_3$ compels $\Delta_{\rm tr}$ to be positive. 
The recent first-principles density functional calculation has estimated that the trigonal crystal field $\Delta_{\rm tr}$ is 
as large as $75$ meV ($\approx 0.27t$) for Na$_2$IrO$_3$~\cite{Foyevtsova2013S}. 
As shown in Fig.~\ref{rixs_dist}, we find that indeed the main peak corresponding to the local $d$-$d$ 
transition is split into double (or multiple) peaks (denoted as `B' and `C' in Fig.~\ref{rixs_dist}) when 
$\Delta_{\rm tr}$ is finite. 
Although a larger $\Delta_{\rm tr}$ seems to be necessary for the double peak structure to 
be clearly visible, our results support the previous studies~\cite{Gretarsson2013S,BHKim2014S} 
in which the origin of the double-peak like structure observed in the RIXS experiment on Na$_2$IrO$_3$ 
is attributed to the local $d$-$d$ excitation in the presence of the trigonal distortion.

\section{Local multiplet structure}

\begin{figure}[b]
\centering
\includegraphics[width=.55\columnwidth]{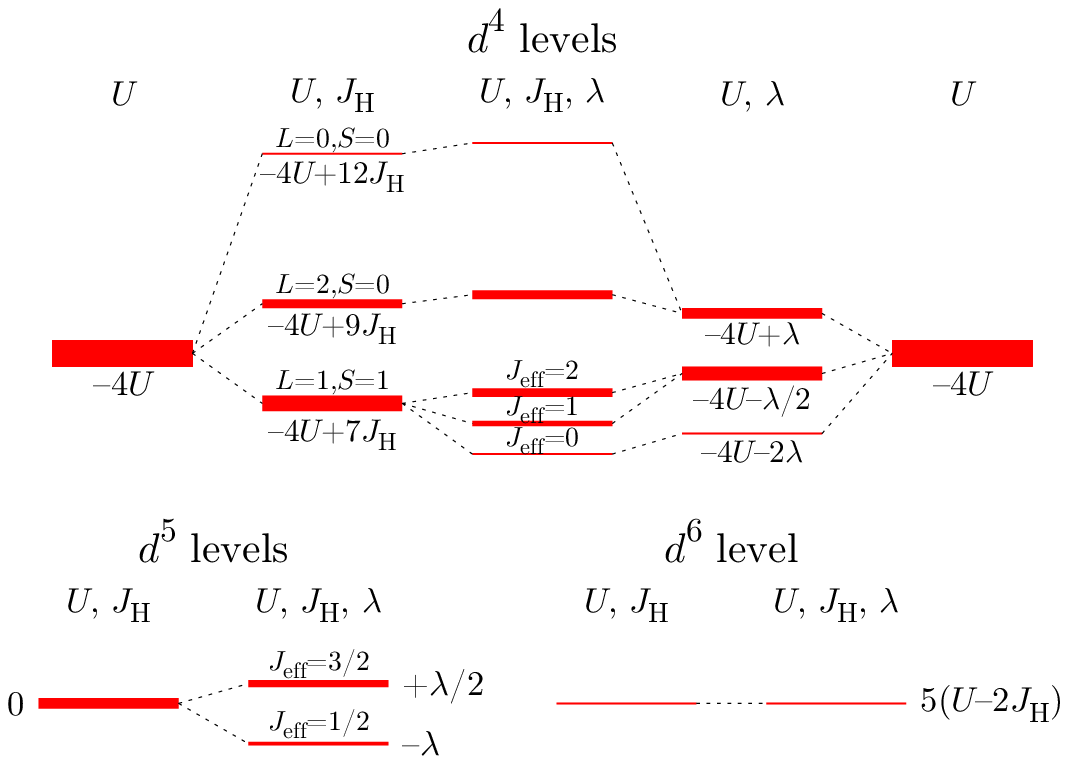}
\caption
{ (Color online)
Schematic energy diagrams of multiplet hierarchy 
in the atomic limit for three relevant electron configurations, i.e., $d^4$, $d^5$,
and $d^6$ in the $t_{2g}$ manifold. 
The multiplet energy levels are represented by red lines (the width is proportional to the degeneracy) when 
non-zero $U$, $J_{\rm H}$, or/and $\lambda$ (indicated at the top of each diagram) are considered. 
The multiplet energy shown by each level is measured from the $d^5$ multiplet energy $E_0$ without 
the SOC. 
}
\label{sfig2}
\end{figure}

It is instructive to examine the energy hierarchy of local multiplets, 
which is obtained simply by solving the single-site terms in Eq.~(\ref{CLH}). 
Figure~\ref{sfig2} summarizes the energy hierarchy of 
three relevant configurations, i.e., $d^4$, $d^5$, and $d^6$ electron configurations in the $t_{2g}$ manifold.

In the case of $d^5$ electron configuration, only the SOC plays a role in determining the doublet ($J_{\rm eff}=1/2$) 
and quartet ($J_{\rm eff}=3/2$) states with their energies $E_0-\lambda$ and $E_0+\frac{\lambda}{2}$, 
respectively, while the Hund's coupling $J_{\rm H}$ cannot have 
any effect on their multiplets. Here, $E_0$ is the energy for $d^5$ electron configuration when $\lambda=0$. 
Because the energy splitting between these two manifolds is $\frac{3}{2}\lambda$, the local $d$-$d$ transition,
which is accessed by the RIXS, appears in this energy region.

On the contrary, in the case of $d^4$ electron configuration, not only $\lambda$ but also $J_{\rm H}$ 
determines the multiplet hierarchy. 
When $J_{\rm H}\ne0$ but $\lambda=0$, fifteen-fold degenerate states split into
nine-fold degenerate states with total angular moment $L=1$ and total spin momentum $S=1$, 
five-fold degenerate states with $L=2$ and $S=0$, and a singlet state with $L=S=0$.  
Their energies are, respectively, $E_0-4U+7J_{\rm H}$, $E_0-4U+9J_{\rm H}$, and $E_0-4U+12J_{\rm H}$, 
thus following the Hund's rule. 
A finite $\lambda$ induces addition splittings. 
The lowest nine-fold degenerate states are split into $J_{\rm eff}=0$, $J_{\rm eff}=1$, and $J_{\rm eff}=2$ states. 
Because there are non-negligible mixing between the $J_{\rm eff}=0$ state and the singlet state with $L=S=0$, 
and also between the $J_{\rm eff}=2$ states and the five-fold degenerate states with $L=2$ and $S=0$, the 
exact energy levels for the $J_{\rm eff}=0$ and $J_{\rm eff}=2$ states are not simply expressed analytically. 
However, the energy difference between the lowest $J_{\rm eff}=0$ and $J_{\rm eff}=2$ states is 
always $\frac{3}{2}\lambda$. 
The energy of the $J_{\rm eff}=1$ states is instead given analytically as $E_0-4U+7J_{\rm H}-\frac{\lambda}{2}$.
When $J_{\rm H}\gg\lambda$, the energies of the lowest $J_{\rm eff}=0$ and $J_{\rm eff}=2$ states are 
$E_0-4U+7J_{\rm H}-\lambda$ and $E_0-4U+7J_{\rm H}+\frac{\lambda}{2}$, respectively. 
In the opposite limit $\lambda\gg J_{\rm H}$, their energies are given as $E_0-4U+7J_{\rm H}-2\lambda$ and
$E_0-4U+7J_{\rm H}-\frac{\lambda}{2}$, respectively.

Note that the optical excitations are determined in the multiplet theory by
the energy difference between the ground state in $d^5d^5$ electron configuration and 
the excited states in $d^4d^6$ electron configuration. 
Therefore, the lowest optical peak appears at $U-3J_{\rm H}$ when $\lambda=0$.
When $\lambda$ is finite, the lowest optical peaks are composed of three subpeaks. 
The central peak, which corresponds to the $J_{\rm eff}=1$ multiplet in the $d^4$ electron configuration, appears at 
$U-3J_{\rm H}+\frac{3\lambda}{2}$. The first (third) lowest peak, related to the $J_{\rm eff}=0$ ($J_{\rm eff}=2$) 
multiplet in the $d^4$ electron configuration, 
appears in an energy range between $U-3J_{\rm H}$ ($U-3J_{\rm H}+\frac{3\lambda}{2}$) and 
$U-3J_{\rm H}+\lambda$ ($U-3J_{\rm H}+\frac{5\lambda}{2}$), 
depending on the relative strength of $J_{\rm H}$ and $\lambda$. 
Since the hopping between the NN $j_{\rm eff}=1/2$ orbitals is absent 
when the hopping is mediated via the edge-sharing oxygens,  
the optical peak attributed to the $J_{\rm eff}=0$ multiplet in the $d^4$ electron configuration is
diminished~\cite{BHKim2014S}.
Contribution of the third multiplet ($J_{\rm eff}=2$) in the $d^4$ electron configuration is
also suppressed because its optical transition amplitude 
between $J_{\rm eff}=1/2$ in the $d^5$ electron configuration and 
$J_{\rm eff}=2$ in the $d^4$ electron configuration is as small as
quarter of that between $J_{\rm eff}=1/2$ in the $d^5$ electron configuration and 
$J_{\rm eff}=1$ in the $d^4$ electron configuration.
Therefore, the honeycomb lattice system in the relativistic Mott insulating limit 
shows a dominant optical peak at 
$U-3J_{\rm H}+\frac{3\lambda}{2}$, attributed to the excitation to the $J_{\rm eff}=1$ multiplet in the 
$d^4$ electron configuration.

\section{Effect of direct hopping $t'$ between neighboring $d$ orbitals}

Even thought the dominant hopping process in the honeycomb lattice is via the ligand $p$ orbitals,
the spatial extension of $4d$ and $5d$ orbitals is large enough to give rise to the direct hopping between 
neighboring TM $d$ orbitals. 
Recent theoretical study based on first-principles density functional calculations has estimated that the 
direct hopping $t'$ between $XY$, $YZ$, and $ZX$ orbitals along the $XY$-, $YZ$-, and $ZX$-directions, 
respectively, 
can be as large as $t'/t\approx 0.10$ for Na$_2$IrO$_3$ and $-0.97$ for $\alpha$-RuCl$_3$~\cite{Winter2016S}.
Therefore, $t'$ is also an important parameter in determining the electronic state specially for $\alpha$-RuCl$_3$.

\begin{figure}[h]
\centering
\includegraphics[width=.65\columnwidth]{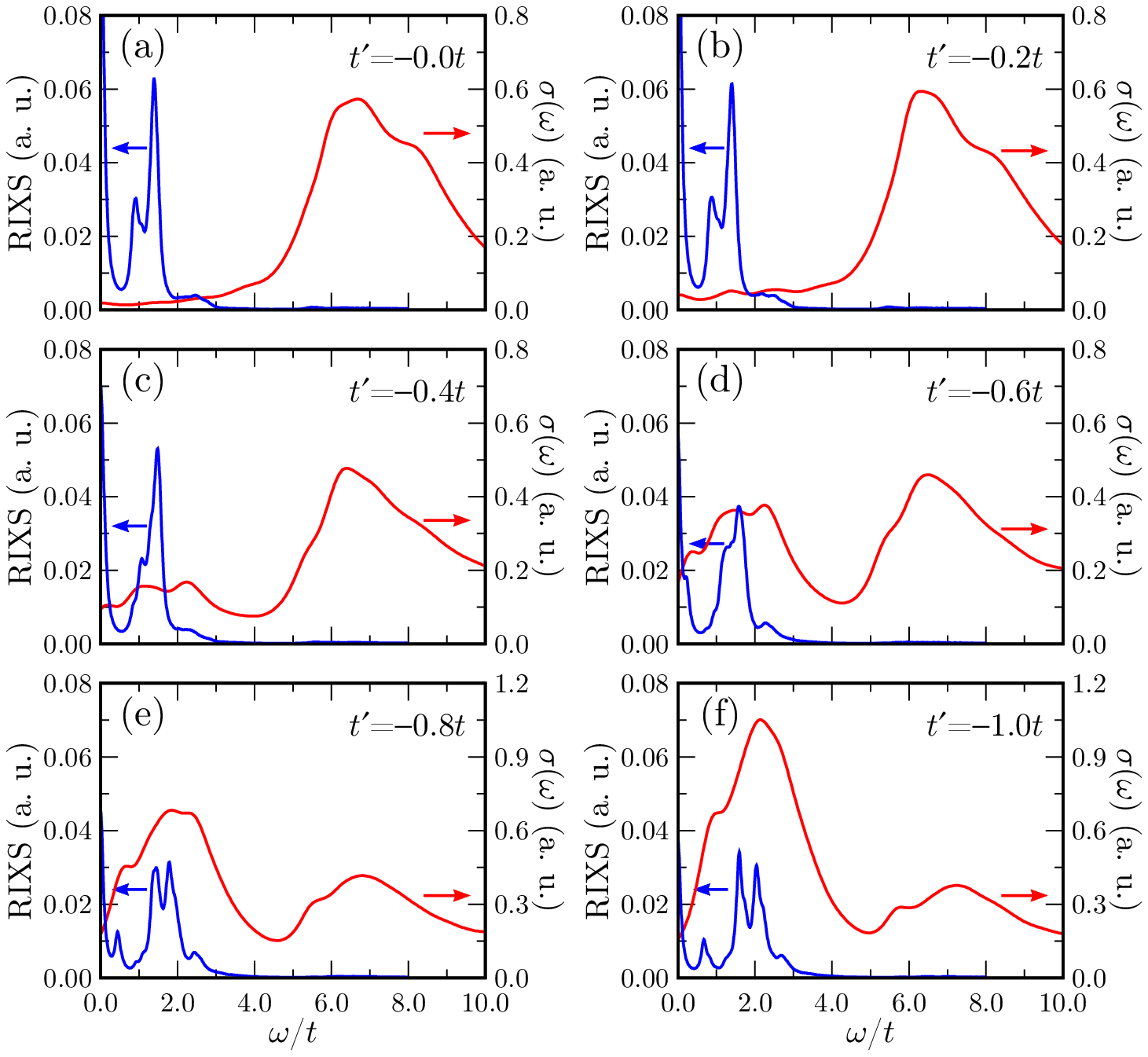}
\caption
{(Color online) 
 OC $\sigma(\omega)$ (red lines) and $L_3$-edge RIXS spectrum at momentum ${\mathbf q}=0$
 (blue lines) for various direct $d$-$d$ hoppings $t'$ 
 indicated in the figures 
 for $U-3J_{\rm H}=6t$, $J_{\rm H}=1.6t$, and $\lambda=0.8t$. 
 Here, no trigonal distortion is considered. 
}
\label{sfig4}
\end{figure}

Figure~\ref{sfig4} presents the OC and RIXS spectra for various values of $t'$ 
when $U-3J_{\rm H}=6t$, $J_{\rm H}=1.6t$, and $\lambda=0.8t$.
An interesting feature is that the low-energy optical peaks arise around $2t$
in addition to the dominant peaks with the large spectral weight near $7t$. 
For small $|t'|\, (\le 0.4t$), 
the small low-energy optical peaks appear near the dominant RIXS peak ($\approx 1.4t$) 
extended up to around 
the higher edge of the RIXS spectrum ($\approx 2.5t$), 
which is in good agreement with the experimental observations~\cite{Plumb2014S,Sandilands2016S}.
Recall that the dominant RIXS peak is attributed to the local $d$-$d$ transition 
from the $j_{\rm eff}=3/2$ to $1/2$ orbitals. 
The direct hopping $t'$ enhances the optical contribution 
of the electron-hole excitation between the neighboring $j_{\rm eff}=1/2$ orbitals,
which is largely suppressed in the edge-shared geometry when $t'=0$.
Our results thus imply that the unusual coupling between these two different types of excitations, 
i.e., the electron-hole excitations between the neighboring $j_{\rm eff}=1/2$ orbitals and the 
local $d$-$d$ excitations,  
can play a role in the low-energy optical excitations. 
This coupling is possible because of the 
finite inter-band transition between the $j_{\rm eff}=1/2$ and $3/2$ bands~\cite{BHKim2014S}. 

For large $|t'|\,(\ge 0.6t)$, we find in Fig.~\ref{sfig4} that a large amount of spectral weights in the OC are 
transferred from the high-energy 
region ($\agt4t$) to the low-energy region ($\alt4t$). 
This observation with a large amount of low-energy spectral weight is far from the 
experimental observations~\cite{Plumb2014S,Sandilands2016S} and thus indicates 
that the first-principles density functional calculations overestimate $|t'|$~\cite{Winter2016S}.


\begin{thebibliography}{99}
\bibitem{hopJ} 
Let $t$ and $t^{\prime}$ be the nearest neighbor hoppings between 
 $YZ$ and $ZX$ orbitals, and between $XY$ orbitals, respectively, 
 along the $XY$-direction. One can show that the nearest neighbor hopping between 
 $j_{\rm eff}=1/2$ orbitals with the same isospin is $t^{\prime}/3$. 
 Therefore, there is no optical transition among the pure $j_{\rm eff}=1/2$ 
 bands when $t^{\prime}=0$.
\end{thebibliography}
\end{document}